\documentclass[twocolumn,nopacs,floatfix,amsmath,nofootinbib,amssymb,preprintnumbers,floatfix]{revtex4}
\usepackage{longtable,lscape,booktabs}
\usepackage{txfonts}
\usepackage{overpic}
\usepackage{amssymb}
\usepackage{indentfirst}
\usepackage{feynmf}
\usepackage{slashed}
\usepackage{cases}
\usepackage{color}
\usepackage{multirow}
\usepackage{ulem}
\usepackage{subfigure}
\usepackage{graphicx,color,dcolumn,bm}
\usepackage[colorlinks,
            citecolor=blue,
            anchorcolor=red,
            menucolor=red,
            linkcolor=red,
            filecolor=red,
            runcolor=red,
            urlcolor=blue,
            frenchlinks=red]{hyperref}

\renewcommand{\arraystretch}{1.50}

\begin{document}

\title{Isospin violation effect and three-body decays of the $T_{cc}^{+}$ state}

\author{Zhi-Feng Sun$^{1,2,3,4}$}\email[]{sunzf@lzu.edu.cn}
\author{Ning Li$^{5}$}\email[]{lining59@mail.sysu.edu.cn}
\author{Xiang Liu$^{1,2,3,4}$}\email[]{xiangliu@lzu.edu.cn}
\affiliation{
$^1$School of Physical Science and Technology, Lanzhou University, Lanzhou 730000, China\\
$^2$Lanzhou Center for Theoretical Physics, Key Laboratory of Theoretical Physics of Gansu Province, Lanzhou University, Lanzhou, Gansu 730000, China\\
$^3$MoE Frontiers Science Center for Rare Isotopes, Lanzhou University, Lanzhou, Gansu 730000, China\\
$^3$Research Center for Hadron and CSR Physics, Lanzhou University and Institute of Modern Physics of CAS, Lanzhou 730000, China\\
$^5$School of Physics, Sun Yat-Sen University, Guangzhou 510275, China}

\date{\today}

\begin{abstract}
In this work, we make a study of $T_{cc}^+$ state observed by the LHCb collaboration in 2021. In obtaining the effective potentials using the One-Boson-Exchange Potential Model we use an exponential form factor, and find that in the short and medium range, the contributions of the $\pi$, $\rho$ and $\omega$ exchanges are comparable while in the long range the pion-exchange contribution is dominant. 
Based on the assumption that $T_{cc}^+$ is a loosely bound state of $D^*D$, we focus on its three-body decay using the meson-exchange method. Considering that the difference between the thresholds of $D^{*+}D^0$ and $D^{*0}D^+$ is even larger than the binding energy of $T_{cc}^+$, the isospin-breaking effect is amplified by the small binding energy of $T_{cc}^+$. Explicitly  including such an isospin-breaking effect we obtain, by solving the Schr\"{o}dinger equation, that the probability of the isoscalar component is about $91\%$ while that of the isovector component is around $9\%$ for $T_{cc}^+$. Using the experimental value of the mass of $T_{cc}^+$ as an input, we obtain the wave function of $T_{cc}^+$ and further obtain its width via the three-body hadronic as well as the radiative decays. The total width we obtain is in agreement with the experimental value of the LHCb measurement with a unitarised Breit-Wigner profile. Conversely, the current results support the conclusion that $T_{cc}^+$ is a hadronic molecule of $D^*D$.
\end{abstract}

\maketitle

\section{Introduction}
The search for the exotic hadrons and the exploration of their nature have been important goals of the particle physics \cite{Brambilla:2019esw,Chen:2016qju,Guo:2017jvc,Liu:2019zoy,Meng:2022ozq,Chen:2022asf,Liu:2024uxn}. The doubly heavy multiquark is particularly important, as its study can deepen our understanding of the nonperturbative strong interaction in the low-energy region and enrich our identification of the hadronic spectrum. In 2021, the LHCb Collaboration announced the observation of the doubly charmed tetraquark $T_{cc}^+$ in the $D^0D^0\pi^+$ mass distribution~\cite{LHCb:2021vvq,LHCb:2021auc}. This discovery is a turning point in hadron physics, providing the first evidence for a clearly exotic meson state with two charm quarks. Using a relativistic P-wave two-body Breit-Wigner function with a Blatt-Weisskopf form factor as the natural resonance profile, the location of the peak relative to the $D^{*+}D^0$ mass threshold, $\delta m_{BW}$, and the width, $\Gamma_{BW}$, are determined to be \cite{LHCb:2021vvq}
\begin{eqnarray}
\delta m_{BW}&=&-273\pm 61\pm 5^{+11}_{-14}\ \text{keV},\nonumber\\
\Gamma_{BW}&=&410 \pm 165 \pm 43^{+18}_{-38}\ \text{keV}.\nonumber
\end{eqnarray}
In order to assess the fundamental properties of near-threshold resonances, the LHCb Collaboration uses advanced parametrisations, i.e., a unitarised Breit-Wigner profile. The extracted mass relative to the $D^{*+}D^0$ threshold and the width are given respectively by \cite{LHCb:2021auc}
\begin{eqnarray}
\delta m_{U}&=&-360\pm 40^{+4}_{-0}\ \text{keV},\nonumber\\
\Gamma_{U}&=&48 \pm 2^{+0}_{-14}\ \text{keV}.\nonumber
\end{eqnarray}

Before this discovery, there were many works searching for the structures with two heavy and two light quarks \cite{Ballot:1983iv,Zouzou:1986qh,Heller:1986bt,Carlson:1987hh,Silvestre-Brac:1992kaa,Manohar:1992nd,Brink:1998as,Cook:2002am,Gelman:2002wf,Janc:2004qn,Vijande:2006jf,Ebert:2007rn,Vijande:2009kj,Yang:2009zzp,Dias:2011mi,Ohkoda:2012hv,Du:2012wp,Li:2012ss,Ikeda:2013vwa,Luo:2017eub,Mehen:2017nrh,Fontoura:2019opw,Xu:2017tsr,Francis:2016hui,Francis:2018jyb,Agaev:2019qqn,Tan:2020ldi,Yang:2019itm,Cheng:2020wxa,Lipkin:1986dw,Navarra:2007yw,Semay:1994ht,Vijande:2007rf,Lee:2009rt,Karliner:2017qjm,Eichten:2017ffp,Wang:2017uld,Junnarkar:2018twb,Liu:2019stu,Ding:2020dio,Molina:2010tx,Pepin:1996id,Vijande:2003ki,Feng:2013kea,Deng:2018kly,Liu:2020nil,Park:2018wjk,Lu:2020rog,Braaten:2020nwp}. The striking feature that $T_{cc}^+$ is extremely close to the threshold. $T_{cc}^+$ has motivated much work investigating this state within the picture of the hadronic molecular state \cite{Chen:2021vhg,Feijoo:2021ppq,Fleming:2021wmk,Meng:2021jnw,Albaladejo:2021vln,Chen:2021tnn,Dai:2021vgf,Deng:2021gnb,Du:2021zzh,Ke:2021rxd,Ling:2021bir,Liu:2019yye,Chen:2021cfl,Dong:2021bvy,Xin:2021wcr,Huang:2021urd,Ren:2021dsi,Zhao:2021cvg,Agaev:2022ast,He:2022rta,Padmanath:2022cvl,Albaladejo:2022sux,Cheng:2022qcm,Abreu:2022sra,Chen:2022vpo,Jia:2022qwr,Wang:2022jop,Dai:2023mxm,Li:2023hpk,Hu:2021gdg,Azizi:2021aib,Abreu:2022lfy,Vidana:2023olz,Dai:2023cyo,He:2023ucd,Jia:2023hvc,Lei:2023ttd,Sakai:2023syt,Montesinos:2023qbx}, tetraquark state \cite{Azizi:2021aib,Abreu:2022lfy,Chen:2021tnn,Agaev:2021vur,Jin:2021cxj,Wu:2022gie,Weng:2021hje,Abreu:2021jwm,Kim:2022mpa,Noh:2023zoq,Liu:2023vrk,Wang:2024vjc}, and many other explanations \cite{Shi:2022slq,Kinugawa:2023fbf,Dai:2023kwv,Ma:2023int,Chen:2021tnn,Yan:2021wdl,Deng:2022cld,Wang:2023iaz}.

In Ref. \cite{Li:2012ss} we investigated the possible tetraquarks with double charm. We found that the interaction between $D^*$ and $D$ is attractive for the system $D^*D$ with quantum number $I(J^P)=0(1^+)$, and it can form a loosely bound state. In the present work, we still consider $T_{cc}^+$ as the $D^{*+}D^0/D^{*0}D^+$ molecular state. Since the mass difference of $D^{*+}D^0$ and $D^{*0}D^+$ is so large compared to the small binding energy of $T_{cc}^+$, the isospin violation effect should not be neglected, and in the present work we  explicitly take this effect into account. 

In the compact tetraquark picture, the heavy diquark-antiquark symmetry contributes a strong attractive force for the doubly heavy systems, so that this scheme can generate a deep bound state relative to the corresponding two-meson threshold. In order to clarify whether the hadronic molecular picture is correct, in this work we develop a method to calculate the three-body decay of $T_{cc}^+$ based on $T_{cc}^+$ being a hadronic molecule of $D^*D$. Note that in Refs. \cite{Meng:2021jnw,Ling:2021bir,Jia:2023hvc}, the authors have already studied the three-body decay of $T_{cc}^+$. Unlike these works, we study the three-body decay of $T_{cc}^+$ in the framework of the $D^{*+}D^0/D^{*0}D^+$ molecular state based on the one-boson-exchange model, where its molecular structure is represented by the wave functions obtained by solving the Schr\"{o}dinger equation.

In this work, we also make an improvement to the one-boson-exchange model. In the past, to calculate the effective potentials a monopole form factor was introduced at each vertex to suppress the high-momentum (or short-range) contribution. However, when the mass of the exchanged particle is large, the monopole form factor does not work very well due to the suppression by its numerator. We therefore choose an exponentially parameterized form factor in the derivation of the effective potential to regulate the short-range interaction.

We also extend the effective Lagrangians from $[\mathrm{SU(3)_L}\otimes \mathrm{SU(3)_R}]_{\mathrm{global}}\otimes [\mathrm{SU(3)_V}]_{\mathrm{local}}$ to $[\mathrm{U(3)_L}\otimes \mathrm{U(3)_R}]_{\mathrm{global}}\otimes [\mathrm{U(3)_V}]_{\mathrm{local}}$, in which $\pi$, $K$, $\eta^{(\prime)}$ are identified as the pseudoscalar Goldstone bosons, while $\rho$, $\omega$, $K^*$ and $\phi$ are taken as the hidden local symmetry gauge bosons. 

The paper is organized as follows. In Sec.~\ref{effective-potential}, we will present the effective potentials with an exponentially parameterized form factor. The formalism for the three-body decay of $T_{cc}^+$ will be discussed explicitly in Sec.~\ref{decay}. The numerical results are then presented and analyzed in Sec.~\ref{numerical-results}. In the final section~\ref{summary} we will summarize our results and draw conclusions. 

%\section{Formalism} \label{formalism}
%\subsection{The one-boson-exchange model with the exponentially parameterized form factor}

\section{Effective Potential}\label{effective-potential}
In this work, we consider the Lagrangians by extending the hidden local symmetry $[\mathrm{SU(3)_L}\otimes \mathrm{SU(3)_R}]_{\mathrm{global}}\otimes [\mathrm{SU(3)_V}]_{\mathrm{local}}$ to $[\mathrm{U(3)_L}\otimes \mathrm{U(3)_R}]_{\mathrm{global}}\otimes [\mathrm{U(3)_V}]_{\mathrm{local}}$. In such case, the exchanged light pseudoscalar and vector mesons can be grouped into compact matrices, respectively, which are as follows
\begin{eqnarray}
M&=&\left(
\begin{array}{ccc}
\frac{\pi^0}{\sqrt{2}}+\frac{\eta}{\sqrt{6}}+\frac{\eta^\prime}{\sqrt{3}}&\pi^+&K^+\\
\pi^-&-\frac{\pi^0}{\sqrt{2}}+\frac{\eta}{\sqrt{6}}+\frac{\eta^\prime}{\sqrt{3}}&K^0\\
K^-&\bar{K}^0&-\frac{2}{\sqrt{6}}\eta+\frac{\eta^\prime}{\sqrt{3}}
\end{array}
\right),\\
\hat{\rho}^\mu&=&\left(
\begin{array}{ccc}
\frac{\rho^0}{\sqrt{2}}+\frac{\omega}{\sqrt{2}}&\rho^+&K^{*+}\\
\rho^-&-\frac{\rho^0}{\sqrt{2}}+\frac{\omega}{\sqrt{2}}&K^{*0}\\
K^{*-}&\bar{K}^{*0}&\phi
\end{array}
\right)^\mu.
\end{eqnarray}
Taking into account also the heavy quark spin symmetry, the Lagrangian of the interaction between the light meson and the heavy meson containing a charm or bottom quark is constructed as \cite{Casalbuoni:1996pg,Casalbuoni:1992gi}
\begin{eqnarray}
\mathcal{L}&=&ig\text{Tr}\left[H^{(Q)}_b\gamma_\mu \gamma_5A^\mu_{ba}\bar{H}^{(Q)}_a\right]\nonumber\\
&&+i\beta\text{Tr}\left[H^{(Q)}_bv_\mu(V^\mu_{ba}-\rho^\mu_{ba})\bar{H}^{(Q)}_a\right]\nonumber\\
&&+i\lambda \text{Tr}\left[H^{(Q)}_b\sigma_{\mu\nu}F^{\mu\nu}_{ba}\bar{H}^{(Q)}_a\right],\label{eq3}
\end{eqnarray}
where
\begin{eqnarray}
H^{(Q)}_a&=&\frac{1+\slashed{v}}{2}\left[P^{*\mu}_a\gamma_\mu-P_a\gamma_5\right],\\
\bar{H}^{(Q)}_a&=&\gamma_0H^{(Q)\dag}_a\gamma_0=\left[P^{*^\dag\mu}_a\gamma_\mu+P^\dag_a\gamma_5\right]\frac{1+\slashed{v}}{2},\\
A^\mu&=&\frac{1}{2}\left(\xi^\dag \partial^\mu \xi-\xi\partial^\mu \xi^\dag\right),\\
V^\mu&=&\frac{1}{2}\left(\xi^\dag \partial^\mu \xi+\xi\partial^\mu \xi^\dag\right),\\
F_{\mu\nu}&=&\partial_\mu \rho_\nu-\partial_\nu \rho_\mu-\left[\rho_\mu,\rho_\nu\right],
\end{eqnarray}
$P=(D^0,D^+,D_s^+)$ and $P^*=(D^{*0},D^{*+},D_s^{*+})$ are the heavy pseudoscalar and vector meson fields, respectively. $\xi=e^{iM/f_\pi}$, and $\rho_\mu=\frac{ig_V}{\sqrt{2}}\hat{\rho}_\mu$. $A^\mu$ is the axial current, while $V^\mu$ is the vector current. In the heavy quark limit, we apply the static limit, i.e., the four-velocity of the heave meson is taken as $v^\mu = (1, 0, 0, 0)$.

After expanding the Lagrangians in Eq.~\eqref{eq3}, we have the terms we need for our calculation
\begin{eqnarray}
\mathcal{L}_{P^{(*)}P^{(*)}M}&=&-i\frac{2g}{f_\pi}\epsilon_{\alpha\mu\nu\lambda}v^\alpha P^{*\mu}\partial^\nu MP^{*\lambda\dag}\nonumber\\
&&-\frac{2g}{f_\pi}(P\partial^\lambda MP_{\lambda}^{*\dag}+P_{\lambda}^*\partial^\lambda MP^\dag),\label{eq1}\\
\mathcal{L}_{P^{(*)}P^{(*)}V}&=&-\sqrt{2}\beta g_V P (v\cdot \hat{\rho})P^\dag\nonumber\\
&&-2\sqrt{2}\lambda g_V\epsilon_{\lambda\mu\alpha\beta}v^\lambda(P\partial^\alpha\hat{\rho}^\beta P^{*\mu\dag}+P^{*\mu}\partial^\alpha\hat{\rho}^\beta P^\dag)\nonumber\\
&&+\sqrt{2}\beta g_VP^{\mu*}(v\cdot \hat{\rho}) P_\mu^{*\dag}\nonumber\\
&&-i2\sqrt{2}\lambda g_VP^{*\mu}(\partial_\mu\hat{\rho}_\nu-\partial_\nu\hat{\rho}_\mu)P^{*\nu\dag}.\label{eq2}
\end{eqnarray}

With the Lagrangians above, we derive the effective potentials, which are related to the scattering amplitudes by 
\begin{eqnarray}
V(\boldsymbol{q})&=&-\frac{\mathcal{M}(\boldsymbol{q})}{4\sqrt{m_1m_2m_3m_4}},
\end{eqnarray}
where $m_i\ (i=1,2,3,4)$ is the mass of the heavy meson in the initial or final state. By performing the Fourier transformation, we get the effective potentials in coordinate space
\begin{eqnarray}
V(r)&=&\frac{1}{(2\pi)^3}\int d^3qe^{i\boldsymbol{q}\cdot \boldsymbol{r}}V(\boldsymbol{q})F^2(\boldsymbol{q}).
\end{eqnarray}
Here, $F(\boldsymbol{q})$ is the form factor, which suppresses the contribution of high momenta, i.e., small distance. And the presence of such a form factor is dictated by the extended (quark) structure of the hadrons \cite{Machleidt:1987hj}. In this work, we adopt the exponentially parameterized form factor 
\begin{eqnarray}
F(\boldsymbol{q})&=&e^{(q_0^2-\boldsymbol{q}^2)/\Lambda^2},
\end{eqnarray}
where $q_0$ is the zero-th component of the four momentum of exchanged meson, and $\Lambda$ is the cutoff. In the past, a monople form factor is usually used, 
\begin{eqnarray}
F_{M}(\boldsymbol{q})&=&\frac{\Lambda^2-m_{ex}^2}{\Lambda^2-q_0^2+\boldsymbol{q}^2}.
\end{eqnarray}
However, in the case of the heavier $\rho$, $\omega$ or $\phi$ exchanges, $F_{M}(\boldsymbol{q})$ is suppressed by the numerator $\Lambda^2-m_{ex}^2$, which may affect the final results. In this work, we will compare the effective potentials and the probability distributions obtained with these two form factors. 

\renewcommand{\arraystretch}{1.5}
\begin{table*}[htbp]
\centering
\caption{The effective potentials.}
\begin{tabular}{c|ccc}\toprule[0.5pt]\toprule[0.5pt]
&$D^{*0}D^+$&$D^{*+}D^0$\\\midrule[0.5pt]
$D^{*0}D^+$&$-V^D_{\rho^0}+V^D_{\omega}+V^C_{\rho^-}+V^C_{\pi^-}$&$2V^D_{\rho^-}-\frac{1}{2}V^C_{\rho^0}+\frac{1}{2}V^C_{\omega}-\frac{1}{2}\tilde{V}^C_{\pi^0}+\frac{1}{6}V^C_{\eta}+\frac{1}{3}V^C_{\eta^\prime}$ \\
$D^{*+}D^0$&$2V^D_{\rho^-}-\frac{1}{2}V^C_{\rho^0}+\frac{1}{2}V^C_{\omega}-\frac{1}{2}\tilde{V}^C_{\pi^0}+\frac{1}{6}V^C_{\eta}+\frac{1}{3}V^C_{\eta^\prime}$&$-V^D_{\rho^0}+V^D_{\omega}+V^C_{\rho^+}+V^C_{\pi^+}$\\\bottomrule[0.5pt]\bottomrule[0.5pt]
\end{tabular}
\label{tab0}
\end{table*}

The expressions of the effective potentials are listed in Table~\ref{tab0} with the definition of the following functions
\begin{eqnarray}
V^D_V&=&\frac{1}{4}\beta^2 g_V^2(\boldsymbol{\epsilon}_1\cdot \boldsymbol{\epsilon}_3^\dag)Y(\Lambda,m_{V},q_0,r),\\
V^C_V&=&2\lambda^2g_V^2\left[\frac{2}{3}\boldsymbol{\epsilon}_1\cdot \boldsymbol{\epsilon}^\dag_4\bigtriangledown^2 Y(\Lambda,\tilde{m}_{V},q_{0},r)\right.\nonumber\\
&&\left.-\frac{1}{3}S(\hat{\boldsymbol{r}},\boldsymbol{\epsilon}_1,\boldsymbol{\epsilon}_4^\dag)r\frac{\partial}{\partial r}\frac{1}{r}\frac{\partial}{\partial r}Y(\Lambda,\tilde{m}_{V},q_{0},r)\right],\\
V^C_p&=&\frac{g^2}{f_\pi^2}\left[\frac{1}{3}\boldsymbol{\epsilon}_1\cdot \boldsymbol{\epsilon}^\dag_4\bigtriangledown^2 Y(\Lambda,\tilde{m}_{p},q_{0},r)\right.\nonumber\\
&&\left.+\frac{1}{3}S(\hat{\boldsymbol{r}},\boldsymbol{\epsilon}_1,\boldsymbol{\epsilon}_4^\dag)r\frac{\partial}{\partial r}\frac{1}{r}\frac{\partial}{\partial r}Y(\Lambda,\tilde{m}_{p},q_{0},r)\right],\\
\tilde{V}^C_p&=&\frac{g^2}{f_\pi^2}\left[\frac{1}{3}\boldsymbol{\epsilon}_1\cdot \boldsymbol{\epsilon}^\dag_4\bigtriangledown^2 U(\Lambda,\tilde{m}_{p}^\prime,q_{0},r)\right.\nonumber\\
&&\left.+\frac{1}{3}S(\hat{\boldsymbol{r}},\boldsymbol{\epsilon}_1,\boldsymbol{\epsilon}_4^\dag)r\frac{\partial}{\partial r}\frac{1}{r}\frac{\partial}{\partial r}U(\Lambda,\tilde{m}_{p}^\prime,q_{0},r)\right].
\end{eqnarray}
In the above, $\tilde{m}_V^2=m_V^2-q_0^2$, $\tilde{m}_p^2=m_p^2-q_0^2$, $\tilde{m}_p^{\prime 2}=q_0^2-m_p^2$ and $q_0=\frac{m^2_2-m^2_1+m^2_3-m^2_4}{2(m_3+m_4)}$.
The functions $Y$ and $U$ are defined as follows 
\begin{eqnarray}
Y(\Lambda,\mu,q_0,r)&=&\int \frac{d^3q}{(2\pi)^3}e^{i\boldsymbol{q}\cdot \boldsymbol{r}}\frac{1}{\boldsymbol{q}^2+\mu^2-i\epsilon}e^{2(q_0^2-\boldsymbol{q}^2)/\Lambda^2}\nonumber\\
&=&-\frac{e^{2q_0^2/\Lambda^2}}{(2\pi)^2r}\frac{\partial}{\partial r}\left\{\frac{\pi}{2\mu}\left[e^{-\mu r}+e^{\mu r}+e^{-\mu r}\right.\right.\nonumber\\
&&\times \text{erf}\left(\frac{r\Lambda}{2\sqrt{2}}-\frac{\sqrt{2}\mu}{\Lambda}\right)- e^{\mu r}\text{erf}\left(\frac{r\Lambda}{2\sqrt{2}}\right.\nonumber\\
&&\left.\left.\left.+\frac{\sqrt{2}\mu }{\Lambda}\right)\right]e^{2\mu^2/\Lambda^2}\right\},\label{eq9}\\
U(\Lambda,\mu,q_0,r)&=&\int \frac{d^3q}{(2\pi)^3}e^{i\boldsymbol{q}\cdot \boldsymbol{r}}\frac{1}{\boldsymbol{q}^2-\mu^2-i\epsilon}e^{2(q_0^2-\boldsymbol{q}^2)/\Lambda^2}\nonumber\\
&=&\frac{e^{2q_0^2/\Lambda^2}}{(2\pi)^2r}\frac{\partial}{\partial r}\left\{\pi\left[-\frac{1}{2i\mu}\left(e^{-i\mu r}\text{erf}\left(\frac{r\Lambda}{2\sqrt{2}}\right.\right.\right.\right.\nonumber\\
&&\left.\left.-\frac{\sqrt{2}i\mu}{\Lambda}\right)-e^{i\mu r}\text{erf}\left(\frac{r\Lambda}{2\sqrt{2}}+\frac{\sqrt{2}i\mu}{\Lambda}\right)\right)\nonumber\\
&&\left.\left.-\frac{i}{\mu}\cos(\mu r)\right]e^{-2\mu^2/\Lambda^2}\right\}.\label{eq10}
\end{eqnarray}
The derivation of Eqs. (\ref{eq9})-(\ref{eq10}) can be found in the Appendix of Ref. \cite{He:2024aej}. 

Using the Gaussian expansion method (GEM) \cite{Hiyama:2003cu}, we solve the coupled-channel Schr\"{o}dinger equation to find the bound state solutions, 
\begin{align}
    \left(\hat{K}+\hat{M}+\hat{V}\right)\Psi=E\Psi.\label{eq66}
\end{align}
Here, $\hat{K}=\text{diag}(-\frac{\tilde{\Delta}}{2\mu_1},-\frac{\tilde{\Delta}}{2\mu_2},\cdots)$, $\hat{M}=\text{diag}(0,M_2-M_1,M_3-M_1,\cdots)$. For the central-force field problem, the full wave function can be decoupled into the radial part and the angular components. After integrating out the angular part, one obtains a one-dimensional radial Sch\"odinger equation
%Schthe system does not depend on the azimuth and polar angle,
where the double derivative operator $\tilde{\Delta}=\frac{1}{r^2}\frac{\partial}{\partial r}\left(r^2\frac{\partial}{\partial r}\right) - \frac{l(l+1)}{r^2}$. For the function $U(\Lambda,\mu,q_0,r)$, which is complex, we use the real part in order to solve the radial part of the stationary Schr\"{o}dinger equation, which is real. 

%\subsection{Three body decay of $T_{cc}^+$}
\section{Three Body Decay of $T_{cc}^+$} \label{decay}

\begin{figure}
\centering
\vspace{0.1cm}
\setlength{\abovecaptionskip}{0cm} 
\includegraphics[width=1.00\linewidth]{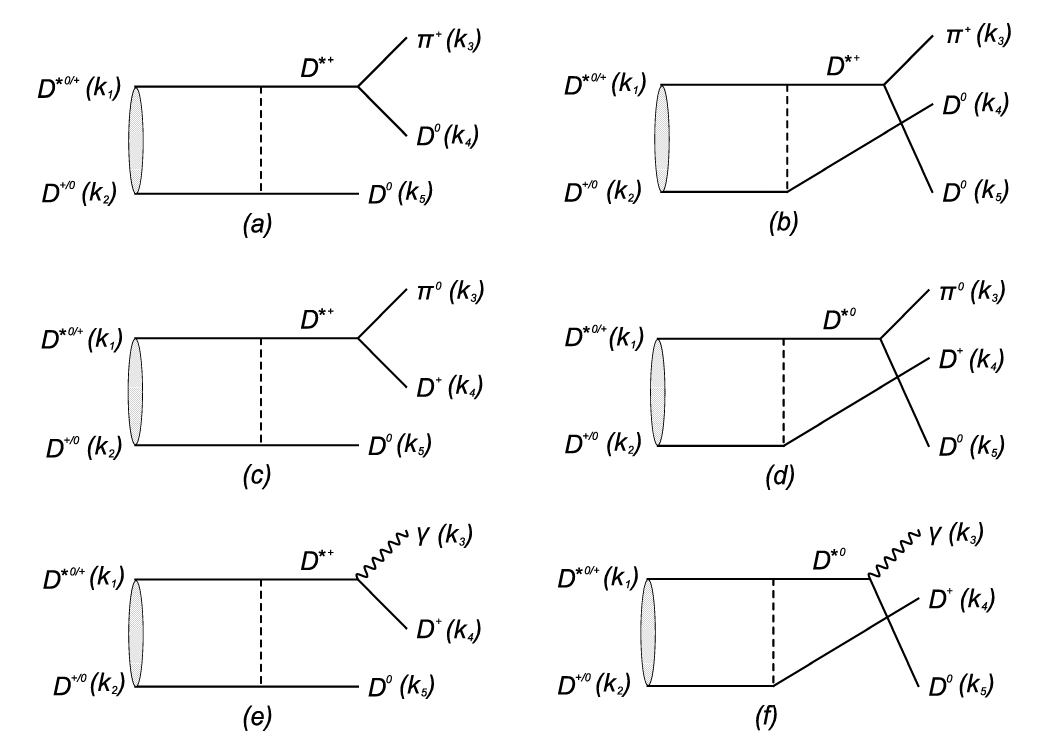}
\caption{The diagrams of the strong and radiative decays of $T_{cc}^+$. The exchanged mesons are $\pi$, $\eta$, $\eta^\prime$, $\rho$ and $\omega$. The same number of diagrams with charmed pseudoscalar (vector) meson connected to vector (pseudoscalar) meson are not shown because their contributions are close to zero, see the main text. \label{fig1}}
\end{figure}

 The main task of the current work is to  derive a practical general formula for the calculation of the strong and radiative three-body decay of $T_{cc}^+$. Here, the $T_{cc}^+$ state is considered as the $D^{*+}D^0/D^{*0}D^+$ molecular state. The decays occur first with the t-channel exchange and then sequentially the vector charmed meson $D^{*}$ decays into $\pi D$ or $\gamma D$. The widths of the strong  and radiative decay are
\begin{eqnarray}
\Gamma_{T_{cc}^{+}\to \pi^{+,0}D^{0,+}D^0} &=&\frac{1}{(2\pi)^3}\frac{1}{32m_{T_{cc}^+}^3}\int dm_{34}^2dm_{45}^2\nonumber\\
&&\times\frac{1}{3}\sum\limits_{S_z^T}|\mathcal{M}_{T_{cc}^+\to \pi^{+,0}D^{0,+}D^0}|^2\frac{1}{S},
\end{eqnarray}
and
\begin{eqnarray}
\Gamma_{T_{cc}^{+}\to \gamma D^{+}D^0} &=&\frac{1}{(2\pi)^3}\frac{1}{32m_{T_{cc}^+}^3}\int dm_{34}^2dm_{45}^2\nonumber\\
&&\times\frac{1}{3}\sum\limits_{S_z^T,S_z^\gamma}|\mathcal{M}_{T_{cc}^+\to \gamma D^{+}D^0}|^2,
\end{eqnarray}
 respectively,
where $S_{z}^T$ and $S_z^\gamma$ are the spin magnetic quantum numbers of $T_{cc}^+$ and the photon respectively. ${1}/{S}$ is the symmetry factor, i.e., ${1}/{2!}$ for $T_{cc}^{+}\to \pi^{+}D^{0}D^0$, $1$ for $T_{cc}^{+}\to \pi^{0}D^{+}D^0$ and  $1$ for $T_{cc}^{+}\to \gamma D^{+}D^0$. $\mathcal{M}_{T_{cc}^+\to \pi^{+}D^{0}D^0}$, $\mathcal{M}_{T_{cc}^+\to \pi^{0}D^{+}D^0}$ and $\mathcal{M}_{T_{cc}^+\to \gamma D^{+}D^0}$ are the amplitudes corresponding to the Feynman diagrams in Fig. \ref{fig1}. In fact, there should be the same number of diagrams with pseudoscalar(vector) charmed meson associated with the vector(pseudoscalar) charmed meson in the t-channel process. However, after  specifically calculating such diagrams, we found that their contributions are close to zero. Therefore, we do not show such diagrams in Fig. \ref{fig1}. 
The Feynmann diagrams Fig. \ref{fig1} (a)-(d) are calculated using the Lagrangians in Eqs. \eqref{eq1}-\eqref{eq2} to obtain the strong decay amplitudes. To calculate the radiative decay amplitude, however, we need two more vertices describing the interactions of $T_{cc}^+$ with $\gamma D^{*0}D^0$ and $\gamma D^{*+}D^+$, i.e.,
\begin{eqnarray}
\mathcal{L}&=&eg_{D^0D^{*0}\gamma}\epsilon^{\mu\nu\alpha\beta}\partial_\mu A_\nu v_\alpha D^{*0}_\beta D^{0\dag},\\
\mathcal{L}&=&eg_{D^+D^{*+}\gamma}\epsilon^{\mu\nu\alpha\beta}\partial_\mu A_\nu v_\alpha D^{*+}_\beta D^{+\dag}.
\end{eqnarray}
The elementary charge $e=0.303$. 
%The coupling constants $g_{D^0D^{*0}\gamma}$ and $g_{D^+D^{*+}\gamma}$, which already involve the higher order corrections studied in Ref. \cite{Stewart:1998ke}. 
To determine the value of the couplings $g_{D^0D^{*0}\gamma}$ and $g_{D^+D^{*+}\gamma}$, we use the decay width of $D^*$ in Ref. \cite{Ling:2021bir}. And we get $g_{D^0D^{*0}\gamma}= 1.911$ GeV$^{-1}$ and $g_{D^+D^{*+}\gamma}= 0.478$ GeV$^{-1}$.

Before calculating the amplitudes corresponding to the diagrams in Fig. \ref{fig1}, we apply a power counting to estimate the contribution of each diagram to simplify the calculation. The order of a quantity $X$ is defined as
$\mathcal{O}(X)=-\log_{10}X$.
Neglecting the higher-order contributions, we obtain the following decay amplitudes 
\begin{eqnarray}
\mathcal{M}_{T_{cc}^+\to \pi^+ D^0D^0}&=&-\sqrt{\frac{2\pi m_{T_{cc}^+}}{E_{D^{*0}}E_{D^+}}}\int_0^\infty dr r j_0(k_5r)u_S^{D^{*0}D^+}(r)\mathcal{A}_{\rho^-}^{(a)}\nonumber\\
&&-\sqrt{\frac{2\pi m_{T_{cc}^+}}{E_{D^{*0}}E_{D^+}}}\int_0^\infty dr r j_0(k_4r)u_S^{D^{*0}D^+}(r)\mathcal{A}_{\rho^-}^{(b)},\nonumber\\\label{eq21}\\
\mathcal{M}_{T_{cc}^+\to \pi^0 D^+D^0}&=&-\sqrt{\frac{2\pi m_{T_{cc}^+}}{E_{D^{*0}}E_{D^+}}}\int_0^\infty dr r j_0(k_5r)u_S^{D^{*0}D^+}(r)\mathcal{A}_{\rho^-}^{(c)}\nonumber\\
&&-\sqrt{\frac{2\pi m_{T_{cc}^+}}{E_{D^{*+}}E_{D^0}}}\int_0^\infty dr r j_0(k_4r)u_S^{D^{*+}D^0}(r)\mathcal{A}_{\rho^+}^{(d)},\nonumber\\\label{eq22}
\end{eqnarray}
\begin{eqnarray}
\mathcal{M}_{T_{cc}^+\to \gamma D^+D^0}&=&-\sqrt{\frac{2\pi m_{T_{cc}^+}}{E_{D^{*0}}E_{D^+}}}\int_0^\infty dr r j_0(k_5r)u_S^{D^{*0}D^+}(r) \mathcal{A}_{\rho^-}^{(e)}\nonumber\\
&&-\sqrt{\frac{2\pi m_{T_{cc}^+}}{E_{D^{*+}}E_{D^0}}}\int_0^\infty dr r j_0(k_4r)u_S^{D^{*+}D^0}(r)\mathcal{A}_{\rho^+}^{(f)}.\nonumber\\\label{eq23}
\end{eqnarray}
Note here that we neglect the D-wave contributions which are much smaller than those of the S-wave. In Eqs. \eqref{eq21}-\eqref{eq23}, the specific expressions of $\mathcal{A}^{(a)}_{\rho^-}, \ldots $   are
\begin{eqnarray}
\mathcal{A}_{\rho^-}^{(a)}&=&i\frac{4g\beta^2g_V^2}{f_\pi}m_{D^0}m_{D^{*+}}\sqrt{m_{D^{*0}}m_{D^+}}\nonumber\\
&&\times \frac{Y(\Lambda,\tilde{M}_{1},\tilde{q}_{1},r)}{k^2-m_{D^{*+}}^2+im_{D^{*+}}\Gamma_{D^{*+}}}\boldsymbol{\epsilon}_1\cdot \boldsymbol{k}_3,\\
\mathcal{A}_{\rho^-}^{(b)}&=&i\frac{4g\beta^2g_V^2}{f_\pi}m_{D^0}m_{D^{*+}}\sqrt{m_{D^{*0}}m_{D^+}}\nonumber\\
&&\times \frac{Y(\Lambda,\tilde{M}_{1},\tilde{q}_{1},r)}{k^{\prime 2}-m_{D^{*+}}^2+im_{D^{*+}}\Gamma_{D^{*+}}}\boldsymbol{\epsilon}_1\cdot \boldsymbol{k}_3,\\
\mathcal{A}_{\rho^-}^{(c)}&=&-i\frac{2\sqrt{2}g\beta^2g_V^2}{f_\pi}m_{D^+}m_{D^{*+}}\sqrt{m_{D^{*0}}m_{D^0}}\nonumber\\
&&\times \frac{Y(\Lambda,\tilde{M}_{1},\tilde{q}_{1},r)}{k^2-m_{D^{*+}}^2+im_{D^{*+}}\Gamma_{D^{*+}}}\boldsymbol{\epsilon}_1\cdot \boldsymbol{k}_3,\\
\mathcal{A}_{\rho^+}^{(d)}&=&i\frac{2\sqrt{2}g\beta^2g_V^2}{f_\pi}m_{D^0}m_{D^{*0}}\sqrt{m_{D^{*+}}m_{D^+}}\nonumber\\
&&\times \frac{Y(\Lambda,\tilde{M}_{2},\tilde{q}_{2},r)}{k^{\prime 2}-m_{D^{*0}}^2+im_{D^{*0}}\Gamma_{D^{*0}}}\boldsymbol{\epsilon}_1\cdot \boldsymbol{k}_3,\\
\mathcal{A}_{\rho^-}^{(e)}&=&-i2eg_{D^+D^{*+}\gamma}\beta^2g_V^2m_{D^+}m_{D^{*+}}\sqrt{m_{D^{*0}}m_{D^0}}\nonumber\\
&&\times \frac{Y(\Lambda,\tilde{M}_{1},\tilde{q}_{1},r)[\boldsymbol{\epsilon}_1\cdot (\boldsymbol{k}_3\times \boldsymbol{\epsilon}_3^\dag)]}{k^2-m_{D^{*+}}^2+im_{D^{*+}}\Gamma_{D^{*+}}},\\
\mathcal{A}_{\rho^+}^{(f)}&=&-i2eg_{D^0D^{*0}\gamma}\beta^2g_V^2m_{D^0}m_{D^{*0}}\sqrt{m_{D^{*+}}m_{D^+}}\nonumber\\
&&\times \frac{Y(\Lambda,\tilde{M}_{2},\tilde{q}_{2},r)[\boldsymbol{\epsilon}_1\cdot (\boldsymbol{k}_3\times \boldsymbol{\epsilon}_3^\dag)]}{k^{\prime 2}-m_{D^{*0}}^2+im_{D^{*0}}\Gamma_{D^{*0}}}.
\end{eqnarray}
Here, $j_0(x)={\sin x}/{x}$ is the spherical Bessel function of order 0. $\tilde{q}_1=\frac{m^2_{D^+}-m^2_{D^{*0}}+m^2_{D^{*+}}-m^2_{D^0}}{2(m_{D^{*+}}+m_{D^0})}$, $\tilde{q}_2=\frac{m^2_{D^0}-m^2_{D^{*+}}+m^2_{D^{*0}}-m^2_{D^+}}{2(m_{D^{*0}}+m_{D^+})}$, $\tilde{M}_1=\sqrt{m_{\rho^-}^2-\tilde{q}_1^2}$, and $\tilde{M}_2=\sqrt{m_{\rho^+}^2-\tilde{q}_2^2}$.
$u_0^{D^{*0}D^+}/r$ and $u_0^{D^{*+}D^0}/r$ are the S-wave radial wave functions of $D^{*0}D^+$ and $D^{*+}D^0$ channels, respectively. $k^\mu=k_3^\mu+k_4^\mu$ is the four-momentum of the intermediate $D^*$ in Fig. \ref{fig1} (a), (c) and (e), while $k^{\prime\mu}=k_3^{\mu}+k_5^{\mu}$ is the four-momentum of the intermediate $D^*$ meson in Fig. \ref{fig1} (b), (d) and (f).

Note that the amplitudes in Eqs. \eqref{eq21}-\eqref{eq23} have the same tensor structure as those in Refs. \cite{Meng:2021jnw,Ling:2021bir}, i.e.,
\begin{eqnarray}
\tilde{\mathcal{M}}_{T_{cc}^+\to \pi^+ D^0D^0}&\sim&k_{\pi^+\mu} \frac{-g^{\mu\nu}+k^\mu k^\nu/m_{D^{*}}^2}{k^2-m_{D^{*}}^2+im_{D^{*}}\Gamma_{D^{*}}}\epsilon_\nu\nonumber\\
&\sim &\boldsymbol{k}_{\pi^+}\cdot {\boldsymbol{\epsilon}},
\end{eqnarray}
\begin{eqnarray}
\tilde{\mathcal{M}}_{T_{cc}^+\to \pi^0 D^+D^0}&\sim&k_{\pi^0\mu} \frac{-g^{\mu\nu}+k^\mu k^\nu/m_{D^{*}}^2}{k^2-m_{D^{*}}^2+im_{D^{*}}\Gamma_{D^{*}}}\epsilon_\nu\nonumber\\
&\sim &\boldsymbol{k}_{\pi^0}\cdot \boldsymbol{\epsilon},
\end{eqnarray}
\begin{eqnarray}
\tilde{\mathcal{M}}_{T_{cc}^+\to \gamma D^+D^0}&\sim&\frac{-g^{\mu\nu}+k^\mu k^\nu/m_{D^{*}}^2}{k^2-m_{D^{*}}^2+im_{D^{*}}\Gamma_{D^{*}}}\epsilon^{\mu\nu\alpha\beta}k_{\gamma\mu}\epsilon_{\gamma\nu} k_\alpha\epsilon_\nu \nonumber\\
&\sim & \boldsymbol{\epsilon}\cdot (\boldsymbol{k}_\gamma\times \boldsymbol{\epsilon}_\gamma).
\end{eqnarray}

\section{Numerical results}\label{numerical-results}

With the analytic results given before, we solve the Schr\"odinger equation to obtain the radial wave function, and then we calculate the decay width based on the radial wave function. In this section we will present the numerical results, perform the discussion and draw the conclusion.  

\subsection{Comparison of the effective potentials with different form factors}

\begin{figure*}
\centering
\vspace{0.5cm}
\setlength{\abovecaptionskip}{0cm} 
\begin{minipage}{7cm}
\includegraphics[width=1.00\linewidth]{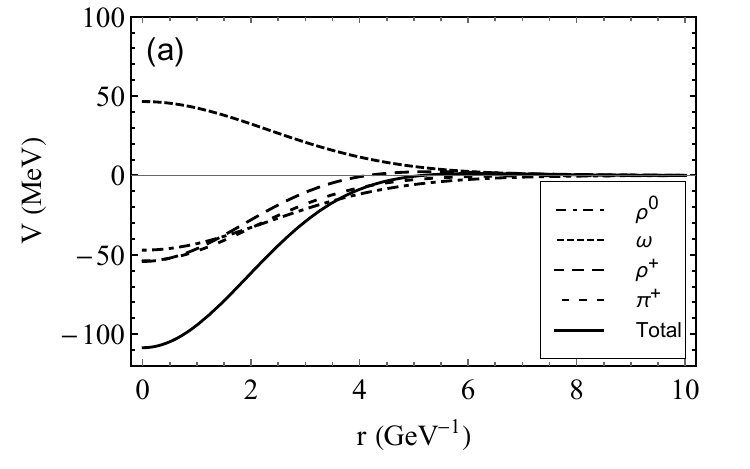}
\end{minipage}
\begin{minipage}{7cm}
\includegraphics[width=1.00\linewidth]{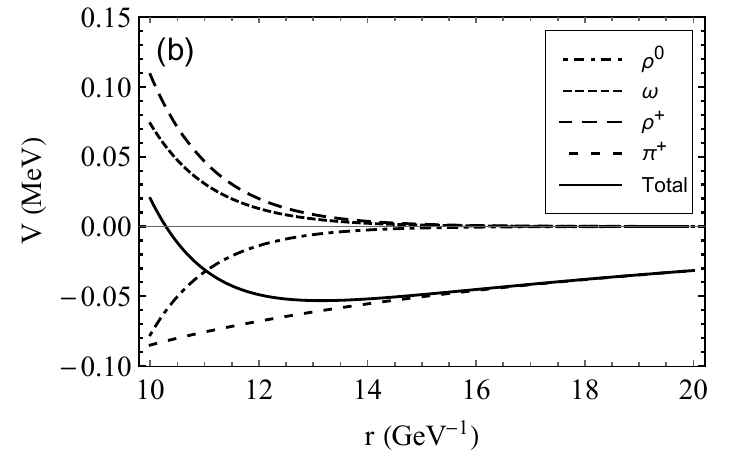}
\end{minipage}
\begin{minipage}{7cm}
\includegraphics[width=0.975\linewidth]{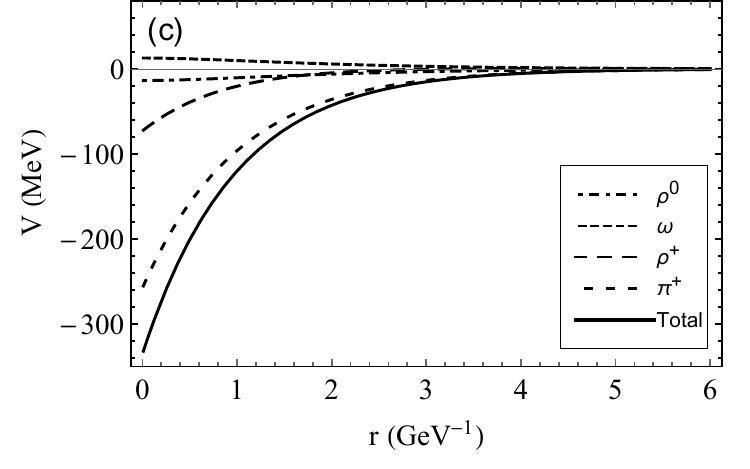}
\end{minipage}
\begin{minipage}{7cm}
\includegraphics[width=0.98\linewidth]{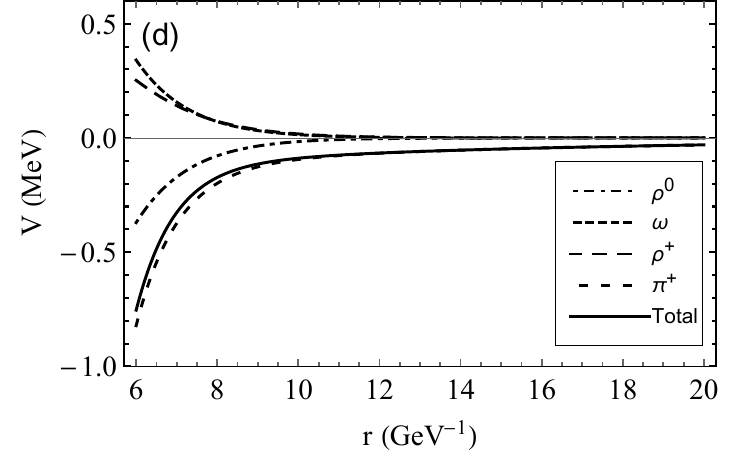}
\end{minipage}
\caption{The effective potentials for the $D^{*+}D^0 \to D^{*+}D^0$ process. (a) and (b) correspond to the exponential form factor with different interaction range, and (c) and (d) the monopole form factor with different interaction range. In all cases, the cutoff $\Lambda$ is chosen to be $1000$ MeV. \label{fig2}}
\end{figure*}

As mentioned above, we use the exponential form factor in our calculation. For comparison, we take the $D^{*+}D^0 \to D^{*+}D^0$ process as an example, and plot the effective potentials with the exponential form factor (upper panel) and the monopole form factor (lower panel) in Fig. \ref{fig2}. It can clearly be seen that for the exponential form factor the pion-exchange contribution and the vector-meson-exchange contribution are comparable in the short and medium interaction range whereas in the long range the pion-exchange contribution is dominant (see Fig. \ref{fig2} (a) and (b)). In the case of the monopole form factor, the vector exchange contribution is relatively small compared to the pion exchange for the whole interaction range. This is because the masses of the vector mesons are larger than that of pion, and their contributions are suppressed by the numerator of the monopole form factor $F_{M}(\boldsymbol{q})=\frac{\Lambda^2-m_{ex}^2}{\Lambda^2-q_0^2+\boldsymbol{q}^2}$, where the cutoff $\Lambda$ is around 1000 MeV.

\subsection{Isospin Symmetry Violation}
The threshold difference betwen $D^{*0}D^+$ and $D^{*+}D^0$ is about $1.34$ MeV, which is much larger than the 
the small binding energy of $T_{cc}^+$. Therefore, the isospin symmetry does not fully hold. Here, we explicitly consider
the isospin symmetry violation effect for $T_{cc}^+$. 

In the following, we apply two types of bases, one is for the different channels (labelled by 1) and the other is for the different isospin states (labelled by 2). The wave functions of $T_{cc}^+$ in these two bases are
\begin{eqnarray}
\psi_{T_{cc}^+}&=&
\left(\begin{array}{c}
\frac{u^{D^{*0}D^+}_S}{r}|^3S_1\rangle\\
\frac{u^{D^{*0}D^+}_D}{r}|^3D_1\rangle\\
\frac{u^{D^{*+}D^0}_S}{r}|^3S_1\rangle\\
\frac{u^{D^{*+}D^0}_D}{r}|^3D_1\rangle
\end{array}\right),\
\psi_{T_{cc}^+}^\prime =
\left(\begin{array}{c}
-\frac{u^{D^{*0}D^+}_S+u^{D^{*+}D^0}_S}{\sqrt{2}r}|^3S_1\rangle\\
-\frac{u^{D^{*0}D^+}_D+u^{D^{*+}D^0}_D}{\sqrt{2}r}|^3D_1\rangle\\
\frac{u^{D^{*0}D^+}_S-u^{D^{*+}D^0}_S}{\sqrt{2}r}|^3S_1\rangle\\
\frac{u^{D^{*0}D^+}_D-u^{D^{*+}D^0}_D}{\sqrt{2}r}|^3D_1\rangle
\end{array}\right),\nonumber
\end{eqnarray}
respectively. These two wave functions are related by each other via the relation $\psi_{T_{cc}^+}^\prime=K\psi_{T_{cc}^+}$ with the transformation matrix 
\begin{eqnarray}
K&=&\left(
\begin{array}{cccc}
-\frac{1}{\sqrt{2}}&0&-\frac{1}{\sqrt{2}}&0\\
0&-\frac{1}{\sqrt{2}}&0&-\frac{1}{\sqrt{2}}\\
\frac{1}{\sqrt{2}}&0&-\frac{1}{\sqrt{2}}&0\\
0&\frac{1}{\sqrt{2}}&0&-\frac{1}{\sqrt{2}}
\end{array}\right).
\end{eqnarray}

\begin{figure*}
\centering
\vspace{0.5cm}
\setlength{\abovecaptionskip}{0cm} 
\begin{minipage}{7.2cm}
\includegraphics[width=1.00\linewidth]{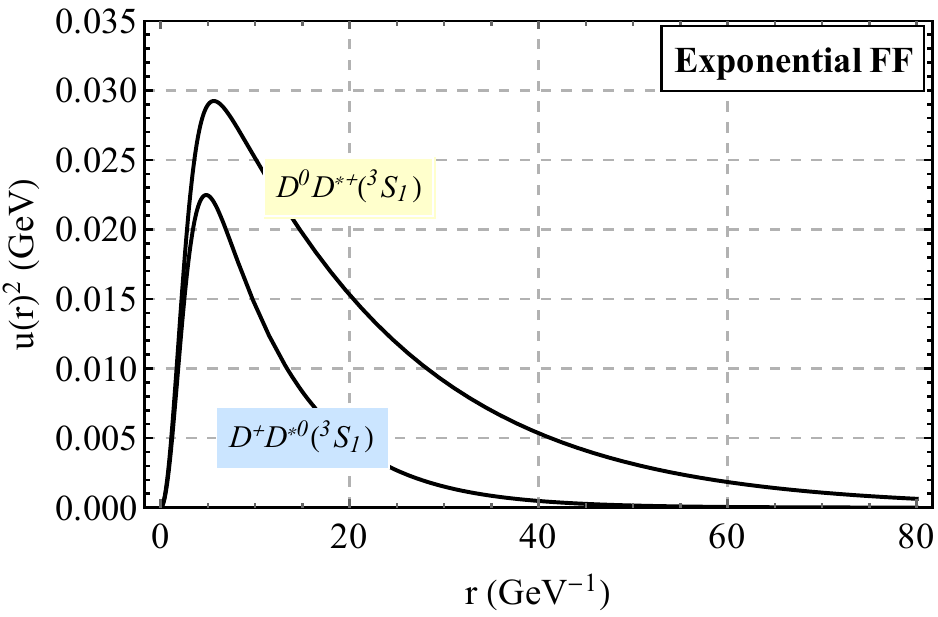}
\end{minipage}
\begin{minipage}{7cm}
\includegraphics[width=1.00\linewidth]{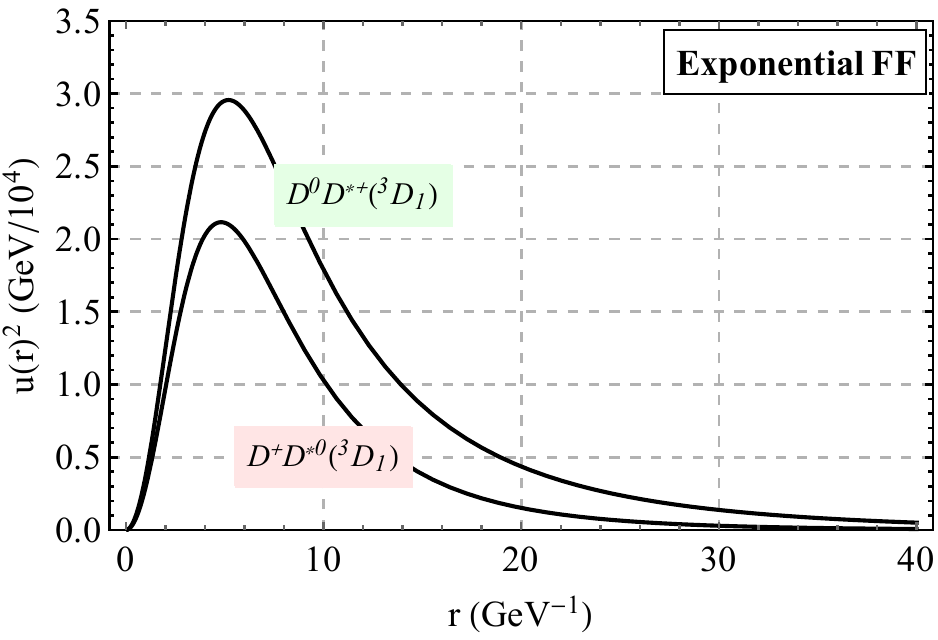}
\end{minipage}
\begin{minipage}{7.3cm}
\includegraphics[width=1.00\linewidth]{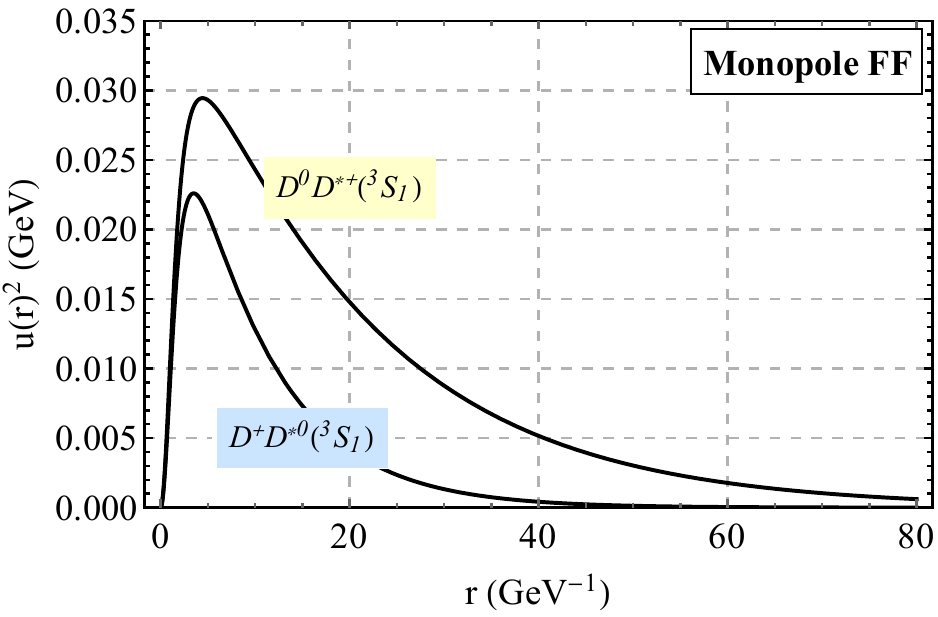}
\end{minipage}
\begin{minipage}{7cm}
\includegraphics[width=1.00\linewidth]{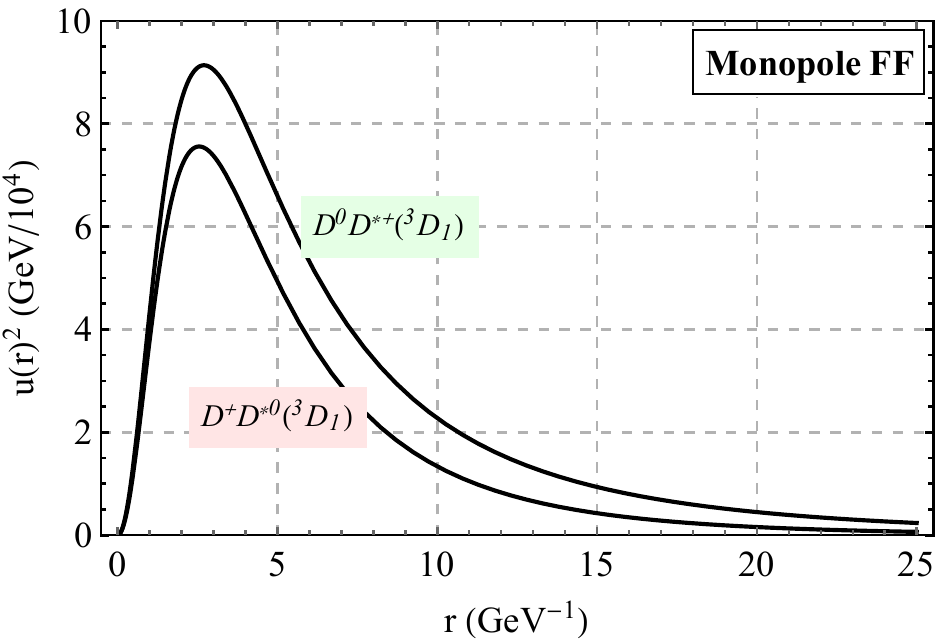}
\end{minipage}
\caption{The radial probability distributions for different channels. (a) and (b) are the results using the exponential form factor with $\Lambda=798$ MeV, while (c) and (d) are the results using the monopole form factor with $\Lambda=1187$ MeV. \label{fig3}}
\end{figure*}

With the effective potentials shown in Sec.~\ref{effective-potential}, we solve the Schr\"{o}dinger equation using the basis $1$. The numerical results are given in Table \ref{tab1}. When the cutoff $\Lambda$ is fixed at 782-798 MeV, we obtain a loosely bound state of $D^*D$ with a binding energy $274-359$ keV, which is in agreement with the experimental value from LHCb. The radial probability distributions of the different channels are shown in Fig. \ref{fig3}, in which the cutoffs are fixed at $798$ MeV and $1187$ MeV for the respective exponential and monopole form factors, so that the binding energies of these two cases are both around $360$ keV. It can clearly be seen that the probability distributions of the S-wave are almost the same. Since the S-wave contributions are dominant, replacing the monopole form factor with the exponential form factor does not change the results too much.
\begin{table}
    \centering
    \caption{The numerical results of the binding energy ($E$), root-mean-square radius ($R_{rms}$) and possibilities ($P$) of different channels. $\Lambda$, $E$, $R_{rms}$ and the probabilities are in units of MeV, keV, fm and $\%$, respectively.}
    \begin{tabular}{cccccccccccc}\toprule[0.5pt]\toprule[0.5pt]
         $\Lambda$&$E$&$R_{rms}$&$P_{D^{*0}D^+(^3S_1)}$&$P_{D^{*0}D^+(^3D_1)}$&$P_{D^{*+}D^0(^3S_1)}$&$P_{D^{*+}D^0(^3D_1)}$\\ \midrule[0.5pt]
         782&200.3&6.7&23.9&0.1&75.7&0.2\\
         786&236.2&6.2&25.3&0.1&74.3&0.2\\
         790&274.5&5.7&26.6&0.1&73.0&0.2\\
         794&315.3&5.4&27.8&0.1&71.9&0.2\\
         798&358.6&5.1&28.9&0.1&70.7&0.2\\ \bottomrule[0.5pt]\bottomrule[0.5pt]           
    \end{tabular}
    \label{tab1}
\end{table}

Considering the basis $2$, the probability of the isovector component within $T_{cc}^+$ is
\begin{eqnarray}
\rho_{10}&=&\int dr\frac{\left[u^{D^{*0}D^+}_S+u^{D^{*+}D^0}_S\right]^2+\left[u^{D^{*0}D^+}_D+u^{D^{*+}D^0}_D\right]^2}{2},
\end{eqnarray}
while that of the isoscalar component is 
\begin{eqnarray}
\rho_{00}&=&\int dr\frac{\left[u^{D^{*0}D^+}_S-u^{D^{*+}D^0}_S\right]^2+\left[u^{D^{*0}D^+}_D-u^{D^{*+}D^0}_D\right]^2}{2}.
\end{eqnarray}
Numerically, if the cutoff is chosen as $790-798$ MeV, the probabilities of $D^{*+}D^0$ and $D^{*0}D^+$ are 73.3\%-71.0\% and 26.7\%-29.0\% respectively. The probability of the isoscalar component is 90.2\%-92.2\% while that of the isovector component is 9.8\%-7.8\%,  revealing a large isospin symmetry violation effect for the loosely bound state of $T_{cc}^+$. 

\subsection{The Three-Body Decay of $T_{cc}^+$}
In Ref. \cite{LHCb:2021vvq}, a relativistic P-wave two-body Breit-Wigner function with a Blatt-Weisskopf form factor is used as the natural resonance profile, while in Ref. \cite{LHCb:2021auc}, a unitarised Breit-Wigner profile is used in the analysis. The experimental widths are $410\pm 165\pm 43^{+18}_{-38}$ keV and $48\pm 2^{+0}_{-14}$ keV, respectively. In this subsection, we will calculate the three-body decay width to investigate which result is reasonable. Using the analytic formulae given in Sec.~\ref{decay} and the radial wave function of $T_{cc}^+$, we calculate the strong and radiative three-body decay of $T_{cc}^+$ and list the numerical results in Table \ref{tab3}.

For strong decays of $T_{cc}^+$, there are two possible channels, i.e., $T_{cc}^+\to D^0D^0\pi^+$ and $T_{cc}^+\to D^0D^+\pi^0$. Note that the decay $T_{cc}^+\to D^+D^+\pi^0$ is kinetically forbidden. If we fix the cutoff $\Lambda$ at $782-798$ MeV, the total width of these two channels is $17.3-22.4$ keV. The partial decay widths of $T_{cc}^+\to D^0D^0\pi^+$ and $T_{cc}^+\to D^0D^+\pi^0$ are $14.8$ keV and $7.6$ keV respectively for the cutoff $\Lambda = 798$ MeV. Note that if the isospin symmetry is exact, in which case the masses of $D^{(*)+}$ and $D^{(*)0}$ are equal, then these two partial widths should be exactly the same. This means that the difference between the partial widths is caused by the isospin symmetry violation effect. 

For radiative decays, the only possible channel is $T_{cc}^+\to D^0D^+\gamma$. If the cutoff is fixed as $782-798$ MeV, the width of the radiative decay is $0.7-1.0$ keV.

Consequently, the total decay width of $T_{cc}^+$ is $23.4$ keV with $\Lambda=798$ MeV, which is close to the lower limit of the experimental value $48\pm 2^{+0}_{-14}$ keV using a unitarised Breit-Wigner profile. This result therefore supports the conclusion that $T_{cc}^+$ is a hadronic molecule composed of $D^{*+}D^0/D^{*0}D^+$.

\begin{table}
    \centering
    \caption{The numerical results of the three-body decay widths as a function of the cutoff. $\Gamma_{R}$, $\Gamma_{S}$ and $\Gamma_{\text{tot}}$ are the radiative, strong and total decay widths.}
    \begin{tabular}{c|ccccccccccc}\toprule[0.5pt]\toprule[0.5pt]
            $\Lambda$ (MeV)     &782  &786  &790  &794  &798 \\ \midrule[0.5pt]
         $\Gamma_{R}$ (keV)     &0.7 &0.8 &0.9 &0.9 &1.0  \\
         $\Gamma_{S}$ (keV)     &17.3&18.6&19.9&21.2&22.4  \\
     $\Gamma_{\text{tot}}$ (keV)&18.0 &19.4 &20.7 &22.1 &23.4 \\ \bottomrule[0.5pt]\bottomrule[0.5pt] 
    \end{tabular}
    \label{tab3}
\end{table}

\section{Summary}\label{summary}
In the present work, we revisit the tetraquark $T_{cc}^+$ observed by the LHCb collaboration in 2021. We consider $T_{cc}^+$ as a $D^{*+}D^0/D^{*0}D^+$ molecular state with the isospin symmetry violation effect explicitly included using the one-boson-exchange potential model. Based on this result, we focus on the three-body strong and radiative decays of $T_{cc}^+$  which can give us much more information about the structure of $T_{cc}^+$. 

In order to calculate the effective potentials, we use the exponential form factor $F(\boldsymbol{q})=e^{(q_0^2-\boldsymbol{q}^2)/\Lambda^2}$ instead of the monopole form factor $F_{M}=\frac{\Lambda^2-m_{ex}^2}{\Lambda^2-q_0^2+\boldsymbol{q}^2}$ which is frequently used in the one-boson-exchange potential model. We find that for the monopole form factor the contribution of $\rho$ and $\omega$ exchanges is much smaller than that of the pion-exchange over the whole interaction range. This is due to the suppression of the numerator of the monopole form factor. However, if we use the exponential form factor, we find a different behavior of the potentials, i.e., in the short and medium interaction range (around $0-0.3$ fm and $0.3-2$ fm, respectively), the vector-meson-exchange contributions are comparable to those of the pion-exchange, while in the long range the pion-exchange potential is dominant. 

In this work, we use the exponential form factor in our calculation. After solving the Schr\"odinger equation, we obtain a loosely bound state of $D^{*+}D^0/D^{*0}D^+$ with binding energy of $358.6$ keV if the cutoff is fixed at $798$ MeV. On the other hand, if we choose a cutoff $790$ MeV, the binding energy is $274.5$ keV. LHCb shows that the binding energy of the $T_{cc}^+$ is $-273\pm 61\pm 5^{+11}_{-14}$ keV and $-360\pm 40^{+4}_{-0}$ keV, when using a relativistic P-wave two-body Breit-Wigner function with a Blatt-Weisskopf form factor as the natural resonance profile and a unitarised Breit-Wigner profile, respectively. Thus, from a spectrum perspective, our results support that $T_{cc}^+$ can be explained as a $D^{*+}D^0/D^{*0}D^+$ molecular state. 

%In addition, we need to emphasize that since the root-mean-square radius of the $T_{cc}^+$ is $5.1-6.7$ fm in which range the pion-exchange potential dominates, that is to say, no matter which of the two form factor we use, the result will not change too much. 

Since the threshold difference between $D^{*+}D^0$ and $D^{*0}D^+$ is large  compared to the small binding energy of $T_{cc}^+$, the isospin symmetry is violated for the $D^{*+}D^0/D^{*0}D^+$ system.  After including an isospin breaking effect, we find that the probability of the isoscalar component is about 91\% and that of the isovector component is about 9\%. 

Since LHCb adopted two methods to analyze the experimental data of the $T_{cc}^+$, there are two different results of its mass and width. In order to check which one is more reasonable, we study the three-body decay of $T_{cc}^+$ and develop for the first time a general method to calculate the three-body decay width. We find that if we take the experimental value of the binding energy, around $360$ keV, as an input  the total width is $23.4$ keV, which is close to the experimental value using a unitarised Breit-Wigner profile. Therefore, our results for the three-body decay width of $T_{cc}^+$ also support that the $T_{cc}^+$ is a hadronic molecule of  $D^{*+}D^0/D^{*0}D^+$.

\section*{Acknowledgments}

We would like to thank Prof. Shi-Lin Zhu for suggestive discussion. This project is supported by the Fundamental Research Funds for the Central Universities under Grant No. lzujbky-2022-sp02, the National Natural Science Foundation of China (NSFC) under Grant Nos. 11705069, 12335001, 11965016 and 12247101, the project for top-notch innovative talents of Gansu province, the National Key Research and Development Program of China under Contract No. 2020YFA0406400, and the Guangdong Basic and Applied Basic Research Foundation under Grant No. 2023A1515011704.

\bibliographystyle{plain}

\end{document}